# V2V COMMUNICATION SURVEY - (WIRELESS TECHNOLOGY)

Mrs. Vaishali D. Khairnar  
Research Scholar  
Institute of Technology  
Nirma University  

Dr. S.N. Pradhan  
Professor  
Institute of Technology  
Nirma University  

**Abstract:** *This paper presents the specific application of wireless communication, **Automotive Wireless Communication** also called as **Vehicle-to-Vehicle Communication**. The paper first gives an introduction to the Automotive Wireless Communication. It explains the technology used for Automotive Wireless Communication along with the various automotive applications relying on wireless communication. Automotive Wireless Communication gives drivers a sixth sense to know what's going on around them to help avoid accidents and improve traffic flow. The paper also describes **VANETS** (vehicular ad hoc networks) and Real-world test network implementation. Finally, the paper is summarized.*

## 1. Introduction:

Using vehicle-to-vehicle (V2V) communication, a vehicle can detect the position and movement of other vehicles up to a quarter of a kilometer away. In a real world where vehicles are equipped with a simple antenna, a computer chip and GPS (Global Positioning System) Technology, your car will know where the other vehicles are, additionally other vehicles will Know where you are too whether it is in blind spots, stopped ahead on the highway but hidden from view, around a blind corner or blocked by other vehicles. The vehicles can anticipate and react to changing driving situations and then instantly warn the drivers with emergency warning messages. If the driver doesn't respond to the alerts message, the vehicle can bring itself to a safe stop, avoiding a collision.

## 2. The V2V Communication:

The radio system for the V2V Communication is derived from the standard IEEE 802.11, also known as Wireless LAN. As soon as two or more vehicles are in radio communication range, they connect automatically and establish an ad hoc network. As the range of a single Wireless LAN link is limited to a few hundred meters, every vehicle is also router and allows sending messages over multi-hop to farther vehicles. The routing algorithm is based on the position of the vehicles and is able to handle fast changes of the ad hoc network topology. In the below picture (fig. 1) we can see that oil is there on the express highway and the a vehicle got slip due to it, this movement is caught by the system and emergency warnings are send to other vehicles about the danger, it even controls the speed of the vehicle to avoid accident.

## 3. Wireless Communication Technologies used for V2V

In this section three Personal Area Network (PAN) standards for in-vehicle communications are presented: Blue-tooth, ZigBee and Ultra Wide Band. Also, one Wireless Local Area Network (WLAN) for inter-vehicle communications is presented: Wi-Fi.

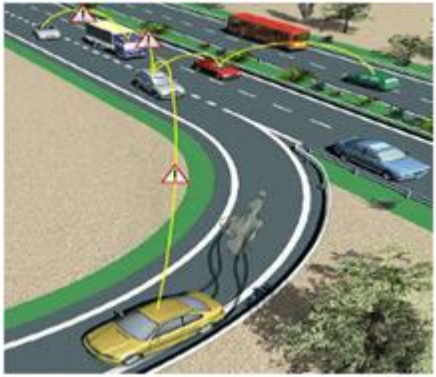

Fig. 1 V2V Communication

### 3.1 Bluetooth (IEEE 802.15.1):

Products that implement the Bluetooth specification can facilitate automatic establishment of a connection between the car's hands-free system (typically part of its audio system) and a mobile phone.

### 3.2 ZigBee (IEEE 802.15.4):

It is a new low-cost, low-power wireless PAN standard, intended to meet the needs of sensors and control devices.

### 3.3 UWB (IEEE 802.15.3a), or Ultra Wide Band:

UWB uses very low-powered, short-pulse radio signals to transfer data over a wide spectrum of frequencies that makes it tolerant to all type of disturbances.

### 3.4 Wi-Fi (wireless fidelity):

This is the general term for any type of IEEE 802.11 network. Examples of 802.11 networks are the 802.11a (up to 54 Mbps), 802.11b (up to 11 Mbps), and 802.11g (up to 54 Mbps). These networks are used as WLANs.

### 4. Vehicular Ad-hoc Networks (VANETs)

Vehicular Ad-hoc Networks (VANETs). VANETs are a special kind of Mobile Adhoc Networks (MANETs). VANET requires fully decentralized network control since no central entity could or should organize the network. Moreover, VANETs hold an additional complexity due to special conditions (i.e., timing and reliability requirements, together with probable saturation when VANET technology is fully deployed).

### 4.1 Interaction with on-board sensors

A VANET platform will most likely include onboard sensors like GPS, to be utilized by network protocols, e.g., by position-based routing. Well-known layered protocol architectures like those in OSI or the Internet have proved to be very useful for traditional (wired) networks. The assumption of protocol layering being the pertinent abstraction leads to the attempt to adapt the traditional protocol stack to the needs of VANETs. In this approach, the principal assignment of functionality to the protocol layers, i.e., covering almost all layers of the OSI reference model remains. Existing layers must be extended by additional functionalities, which are specific to VANETs.VANET applications will most likely evaluate the information contained in a packet, merge it with their own state and then decide how to communicate this updated information. This operation is known as 'in-network' processing. (fig.2)

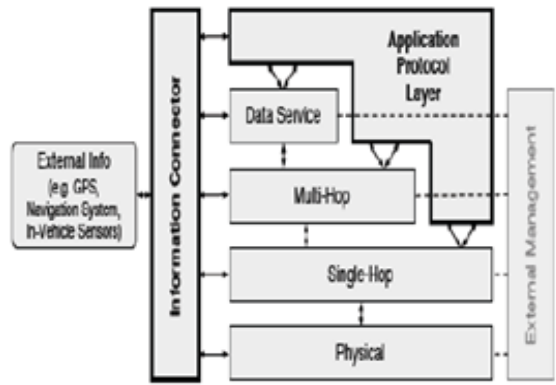

Fig. 2 VANET Architecture

*Information connector*: All protocols can connect to an 'information connector', i.e., a common interface that efficiently exchanges sensor update information, data extracted from packets, and state information (and their change) of protocol layers and devices.

*External management plane*: The external management plane symbolizes a configuration interface to set long-term system settings. In the sense of this proposal, it is not involved in the dynamic self-organization motivated by the different network conditions.

**The *single-hop layer*** incorporates all functionality dealing with communication to direct radio neighbors.

**The *multi-hop layer*** contains protocol elements for forwarding packets to non-neighbored nodes, using neighbors as forwarders.

**The *data service layer*** represents the rest between multihop packet forwarding and the application.

**5 Real-time issues**

Navigation and traffic information systems require position and Internet-like communications, providing traffic information and directions. Voice applications have slightly higher requirements, e.g., real-time voice processing and recognition. However, some safety-systems do have real-time requirements, e.g., communications between the vehicle and other vehicles or roadside objects, implementing collision detection/avoidance systems or
active suspension systems that respond to highway road conditions

**6. GPS Tracking V2V System:**

GPS Tracking V2V is a powerful, high security and high performance vehicle-to-vehicle GPS tracking system. It uses a tablet PC or Laptop plus a RF or GSM/GPRS modem to send, receive and display location data of vehicles of a small group. It is designed for special applications such as police vehicle tracking within a mission group etc.

**8. Applications and Benefits:**
• Navigation and traffic information systems - A vehicle equipped with a telematics unit can direct a driver to a desired location, while providing real-time traffic information.
• Voice recognition and wireless Internet connection –Drivers and their passengers can receive and send voice-activated e-mails while on the road.
• Safety systems - Collision avoidance systems, unsafe driving profiling, intelligent airbag deployment systems, communication between the vehicle and roadside objects. Automatic airbag deployment notification. Accident and roadside assistance. The General Motor's Advanced Automatic Crash Notification system available.
• Security systems - Vehicle antitheft and stolen vehicle tracking services. On
equipped vehicles provide tracking and remote door unlocking.
• Diagnostics and maintenance services - Remote diagnostics and/or maintenance systems, vehicle and driver monitoring.

## 9. Summary

There are several open issues to be addressed. First, which wireless automotive applications rely on real-time systems and how existing research on wireless real-time communications can provide support for these applications. Other important points are related with the integration of wireless networking technologies and with the interoperability problems, which could be expected in the automotive domain. Finally, it should be discussed how these wireless technologies should be integrated in the existing communications architecture comprising several network protocols

**References:**
[1] Mobile communication: by Jochen Schiller.
[2] Car2Car Communication Consortium.
http://www.car-2-car.org/.
[3] IEEE 802.11, The Working Group Setting the Standards for Wireless LANs.
http://www.ieee802.org/11/.
[4] Ultra wide band planet. com.
http://www.ultrawidebandplanet.com/.
[5] ZigBee Alliance.
http://www.zigbee.org/.